\documentclass[sigconf]{acmart}

\usepackage{xspace}  

\newcommand{\eg}{\emph{e.g.},\xspace}
\usepackage{multirow}
\usepackage{bm}
\usepackage{graphicx}
\usepackage{caption}
\usepackage{subcaption}
\usepackage{booktabs}
\usepackage{adjustbox}
\usepackage{float}
\usepackage{color}
\usepackage{longtable}
\usepackage{tcolorbox}
\usepackage{cuted}
\usepackage{listings}
\usepackage{xcolor}
\usepackage{balance}

\AtBeginDocument{%
  }

\copyrightyear{2025}
\acmYear{2025}
\setcopyright{acmlicensed}\acmConference[MM '25]{Proceedings of the 33rd ACM International Conference on Multimedia}{October 27--31, 2025}{Dublin, Ireland}
\acmBooktitle{Proceedings of the 33rd ACM International Conference on Multimedia (MM '25), October 27--31, 2025, Dublin, Ireland}
\acmDOI{10.1145/3746027.3758225}
\acmISBN{979-8-4007-2035-2/2025/10}

\begin{document}

\title{MMESGBench: Pioneering Multimodal Understanding and Complex Reasoning Benchmark for ESG Tasks}

\author{Lei Zhang}
\orcid{0000-0002-5808-5313}
\affiliation{%
  \institution{Nanyang Technological University}
  \city{Singapore}
  \country{Singapore}
}
\email{leizhanzzl.1103@gmail.com}

\author{Xin Zhou}
\orcid{0000-0003-0948-8033}
\affiliation{%
  \institution{Nanyang Technological University}
  \city{Singapore}
  \country{Singapore}
}

\author{Chaoyue He}
\orcid{0000-0002-5728-9672}
\affiliation{%
  \institution{Nanyang Technological University}
  \city{Singapore}
  \country{Singapore}
}

\author{Di Wang}
\orcid{0000-0002-3171-4001}
\affiliation{%
  \institution{Nanyang Technological University}
  \city{Singapore}
  \country{Singapore}
}

\author{Yi Wu}
\orcid{0000-0002-9867-6898}
\affiliation{%
  \institution{University College London}
  \city{London}
  \country{United Kingdom}
}

\author{Hong Xu}
\orcid{0000-0003-1389-5408}
\affiliation{%
  \institution{Nanyang Technological University}
  \city{Singapore}
  \country{Singapore}
}

\author{Wei Liu}
\orcid{0000-0003-3646-3456}
\affiliation{%
  \institution{Alibaba Group}
  \city{Hangzhou}
  \country{China}
}

\author{Chunyan Miao}
\authornote{Corresponding author.}
\orcid{0000-0002-0300-3448}
\affiliation{%
  \institution{Nanyang Technological University}
  \city{Singapore}
  \country{Singapore}
}
\email{ascymiao@ntu.edu.sg}

\renewcommand{\shortauthors}{Lei Zhang et al.}

\begin{abstract}
Environmental, Social, and Governance (ESG) reports are essential for assessing sustainability, regulatory compliance, and financial transparency. However, these documents are typically long, multimodal, and structurally complex, combining dense text, tables, figures, and layout-sensitive semantics. Existing AI systems often struggle to perform reliable document-level reasoning in such settings, and no dedicated benchmark currently exists in ESG domain. To fill the gap, we introduce \textbf{MMESGBench}, a first-of-its-kind benchmark dataset targeted to evaluate multimodal understanding and reasoning across multi-source ESG documents. This dataset is constructed via a human-AI collaborative, multi-stage pipeline. First, a multimodal LLM generates candidate question-answer (QA) pairs by jointly interpreting textual, tabular, and visual information from layout-aware document pages. Second, an LLM verifies the semantic accuracy, completeness, and reasoning complexity of each QA pair. This automated process is followed by an expert-in-the-loop validation, where domain specialists validate and calibrate QA pairs to ensure quality, relevance, and diversity. MMESGBench comprises 933 validated QA pairs derived from 45 ESG documents, spanning across seven distinct document types and three major ESG source categories. Questions are categorized as single-page, cross-page, or unanswerable, with each accompanied by fine-grained multimodal evidence. Initial experiments validate that multimodal and retrieval-augmented models substantially outperform text-only baselines. MMESGBench is publicly available as an open-source dataset at https://github.com/Zhanglei1103/MMESGBench.
\end{abstract}




\maketitle

\section{Introduction}
Understanding Environmental, Social, and Governance (ESG) factors has become increasingly critical in driving sustainable development, responsible investment, and global regulatory compliance~\cite{krabbe2015aligning, xiao2025impact}. This trend has spurred the proliferation of standardized ESG reporting, yielding vast quantities of semantically rich, multimodal data that require interpretation by diverse stakeholders, including governments, corporates, and financial institutions.
Concurrently, advancements in large language models (LLMs) offer considerable potential for automating complex ESG-related tasks such as scoring, investment analysis, risk detection, and compliance monitoring~\cite{zou2025esgreveal,zhang2025optimizing,wu2024susgen}. However, the development and deployment of such automated solutions, which depend on reliable multimodal reasoning, are currently impeded by the scarcity of high-quality, task-specific datasets and benchmarks.

One of the fundamental challenges in advancing AI for ESG applications lies in the inherent complexity of ESG documents, which can be summarized in the fellowing three interrelated dimensions. 1)\textbf{ Multi-source}: ESG documents originate from a broad array of sources, including corporate ESG reports (\eg annual ESG reports and CDP Climate Responses), ESG standards and frameworks (\eg GRI~\cite{gri2023standards}, TCFD~\cite{fsb2017tcfd}, SASB~\cite{sasb2023standards}), and publications from governments and international organizations (\eg IPCC~\cite{ipcc2023synthesis}, SDGs~\cite{un2015agenda2030}). This diversity leads to substantial heterogeneity in document formats, structures, and semantic conventions. 2) \textbf{Multimodality}: ESG reports contain rich multimodal content, integrating narrative text, tabular data, analytical charts, visual figures, and layout-aware cues. Effective understanding requires models to jointly process and reason across these heterogeneous modalities. 3) \textbf{Structural complexity}: ESG documents often span hundreds to thousands of pages and exhibit nested structures, inter-referenced sections, and long-range dependencies, posing challenges for both local comprehension and global document-level reasoning. Despite the increasing demand for automated ESG analysis, existing LLM-based systems remain inadequate for such settings. To the best of our knowledge, no current benchmark captures the full extent of the multimodal and structural reasoning challenges posed by ESG documents.

To address these challenges, we introduce \textbf{MMESGBench}, the first multimodal benchmark designed for ESG document understanding and reasoning. MMESGBench is constructed through a human-AI collaborative pipeline that combines automatic generation, model-based validation, and expert refinement. Specifically, a multimodal LLM first generates candidate question-answer (QA) pairs by jointly interpreting textual content, tables, figures, and layout cues. These pairs are then filtered by an LLM verifier, which checks for factual correctness, completeness, and reasoning validity. Finally, ESG and AI experts review and refine the examples to ensure quality and domain relevance. This process enables scalable generation of evidence-grounded QA data with high fidelity. MMESGBench comprises 933 validated QA pairs across 45 ESG documents, covering seven document types and three major ESG source categories. Each QA pair is labeled as either single-page, cross-page, or unanswerable, and is associated with detailed multimodal evidence, reflecting realistic document-level reasoning scenarios.

MMESGBench is poised to benefit diverse applications within the ESG realm, ranging from ESG document understanding and reasoning to the fine-tuning of specialized ESG models and the implementation of advanced retrieval-augmented generation (RAG) methodologies.
To showcase the capabilities of MMESGBench, we benchmark the ESG document reasoning performance of multiple representative models, including text-only LLMs, multimodal LLMs, and RAG pipelines. 
Results show that multimodal LLMs consistently outperform text-only models, particularly on layout-sensitive and visually grounded questions. In addition, RAG-based methods substantially improve performance on cross-page reasoning tasks by enhancing long-range information access. These findings highlight the importance of multimodal integration and retrieval-aware modeling for reliable ESG document understanding. Thus, MMESGBench is curated as a challenging and practical testbed for future research in this domain.
Our contributions are: 
\begin{itemize}
    \item We release \textbf{MMESGBench}, the first dataset for multimodal QA over real-world ESG documents, covering diverse source types, formats, and modalities.
    \item We propose a \textbf{quality-controlled human-AI collaborative} pipeline, enabling the creation of accurate, diverse, and evidence-grounded QA pairs at scale.
    \item We establish a \textbf{comprehensive benchmark} with strong baseline results and actionable insights, enabling future research on multimodal document-level understanding.
\end{itemize}

\begin{figure*}[t]
    \centering
    \includegraphics[width=0.95\linewidth]{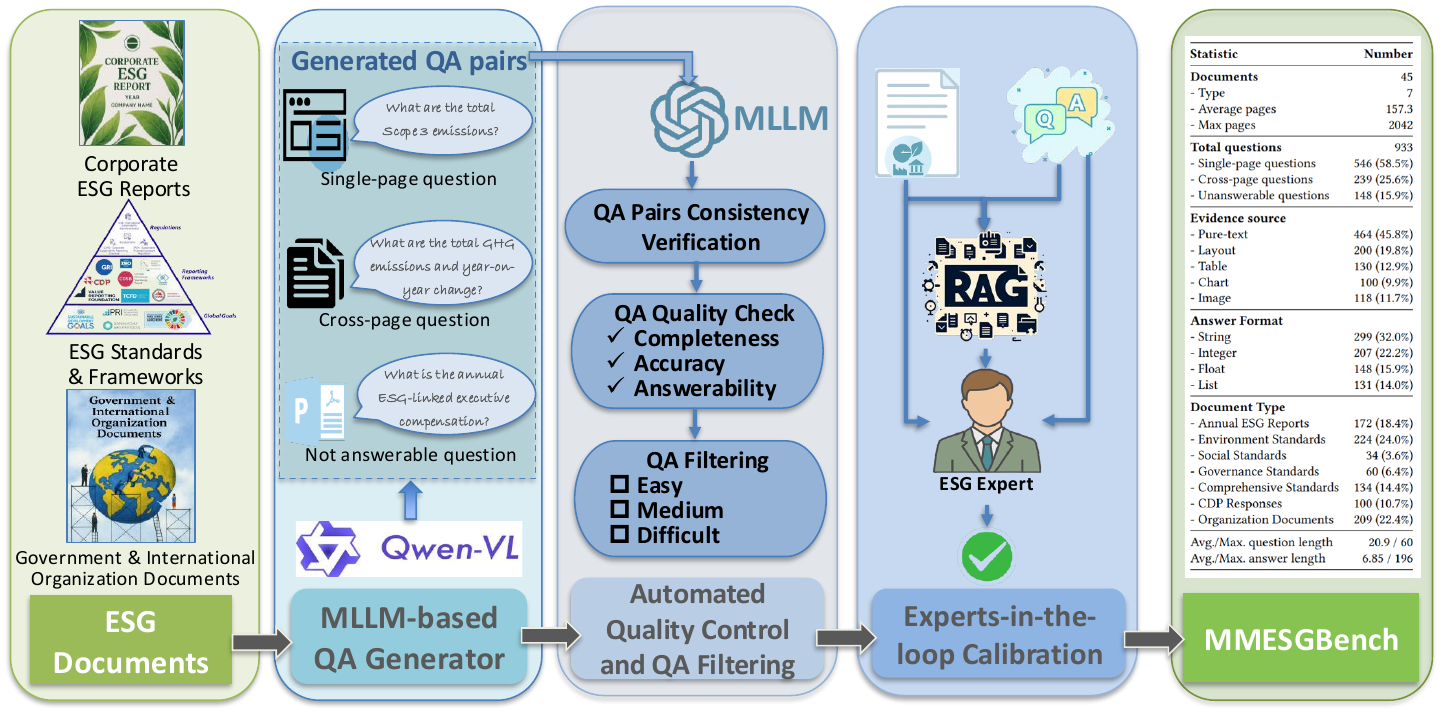}
    \caption{The human-AI collaborative multi-stage QA generation framework.}
    \label{fig:pipeline}
\end{figure*}

\section{Related Work}
\subsection{Document Understanding Benchmarks}
Document understanding is a key research area in multimodal AI. Early efforts such as DocVQA~\cite{mathew2021docvqa} and InfographicsVQA~\cite{mathew2022infographicvqa} focus on single-page visual QA, while TableQA~\cite{sun2020tableqa}, ChartQA~\cite{masry2022chartqa}, and PlotQA~\cite{methani2020plotqa} address isolated modalities such as tables or charts. MP-DocVQA~\cite{tito2023hierarchical} extends DocVQA to multi-page documents but lacks cross-page questions. DUDE~\cite{van2023document} introduces a small number of cross-page questions, with short documents and crowd-sourced annotations. SlideVQA~\cite{tanaka2023slidevqa} includes longer documents (around 20 pages) and cross-page questions but is limited by the slide-deck format and sparsity in reasoning graphs. FinanceBench~\cite{islam2023financebench} offers expert-designed QA over financial reports, but its open-ended format requires manual evaluation. ESGenius~\cite{he2025esgenius} focuses on text-only ESG QA, without incorporating multimodal information. Recent datasets like MMLongBench~\cite{ma2025mmlongbench} and LongDocURL~\cite{deng2024longdocurl} explore long-document multimodal QA and retrieval over web and academic documents, focusing on page-level localization and reasoning. {While these datasets advance document-level QA, they fall short in handling extremely long documents with complex structures, such as ESG materials, which are inherently multi-source, multimodal, and structurally complex. To the best of our knowledge, there currently exists no benchmark for multimodal ESG document understanding, despite its rising importance and complexity.}

\begin{figure*}[t]
  \centering
  \begin{minipage}[b]{0.33\linewidth}
    \centering
    \includegraphics[width=\linewidth]{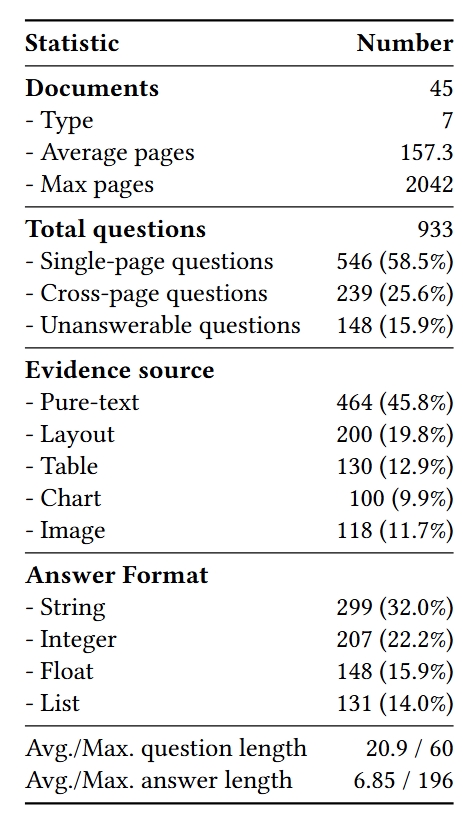}
    \subcaption{Dataset statistics}
  \end{minipage}
  \hfill
  \begin{minipage}[b]{0.64\linewidth}
    \centering
    \begin{minipage}[b]{\linewidth}
      \begin{subfigure}[b]{0.62\linewidth}
        \centering
        \includegraphics[width=\linewidth]{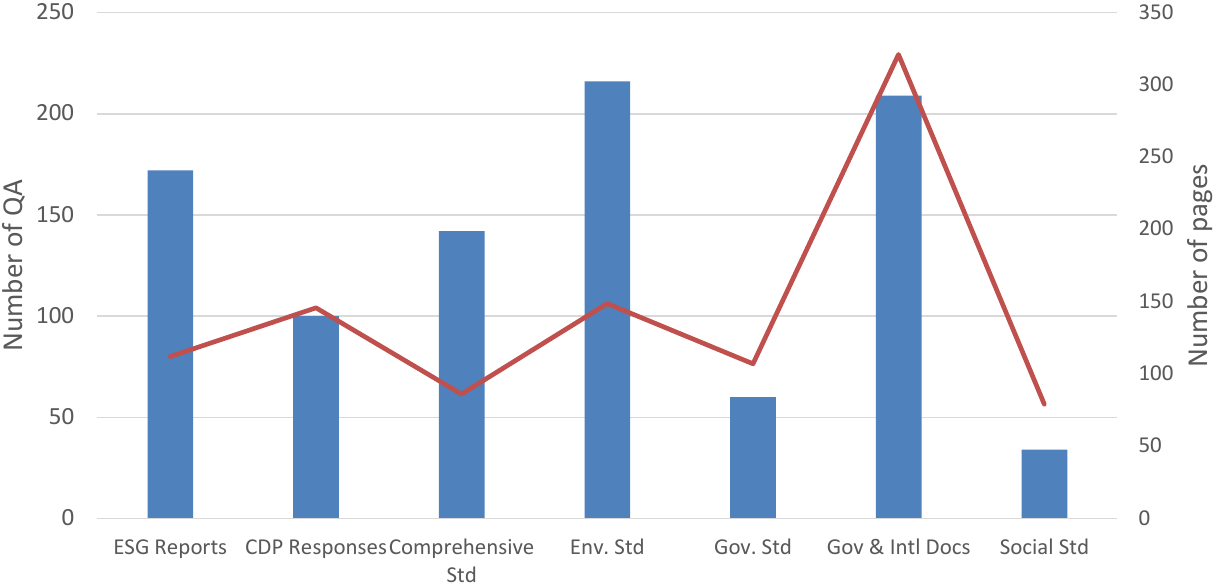}
        \subcaption{QA and pages distribution}
        \label{fig:qa-vs-pages}
      \end{subfigure}
      \hfill
      \begin{subfigure}[b]{0.34\linewidth}
        \centering
        \includegraphics[width=\linewidth]{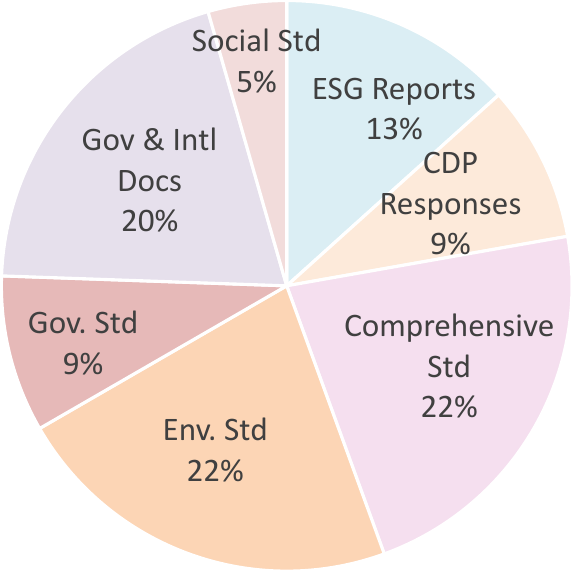}
        \subcaption{Document type breakdown}
        \label{fig:doc-types}
      \end{subfigure}
    \end{minipage}

    \vspace{0.8em}

    \begin{subfigure}[b]{\linewidth}
      \centering
      \includegraphics[width=\linewidth]{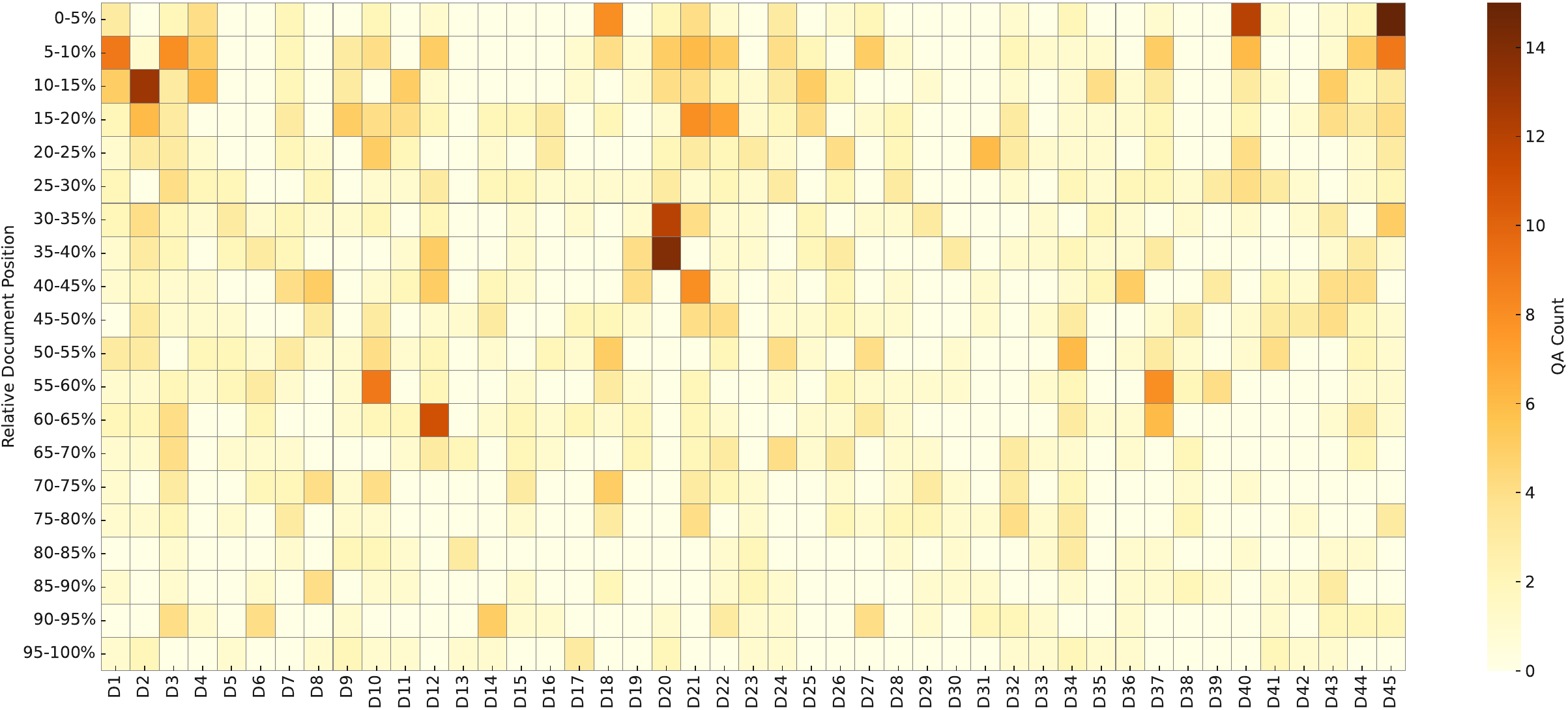}
      \subcaption{Evidence localization heatmap}
      \label{fig:heatmap}
    \end{subfigure}
  \end{minipage}

  \caption{Overview of MMESGBench. (a) Dataset statistics. (b) Distribution of QA instances and total pages per document type; (c) Proportional breakdown of document types; (d) Evidence density heatmap by document and structural position (5\% intervals)}
  \label{fig:dataset-overview}
\end{figure*}

\subsection{AI Applications in ESG Domain}
Advances in NLP and LLMs have driven interest in applying AI to ESG analysis, addressing scale, heterogeneity, and regulatory alignment in disclosures~\cite{arslan2024sustainable,thulke2024climategpt,LeeKJ25,zhou2024advancing}. Prior work such as ESGReveal~\cite{zou2025esgreveal} proposes LLM-based pipelines for structured data extraction from ESG reports, enabling downstream automation of compliance and analytics. E-BERT~\cite{zhang2025optimizing} and ESG-KIBERT~\cite{LeeKJ25} extend pre-trained transformer models with domain-specific fine-tuning, incorporating ESG knowledge and taxonomy constraints to improve entity recognition across standards like GRI and SASB. For retrieval-augmented tasks, ESG-CID~\cite{ahmed2025enhancing} introduces a disclosure index aligned with ESRS-GRI frameworks, improving semantic matching and information access. In parallel, multi-source applications such as knowledge graph-augmented QA~\cite{gupta2024knowledge} over ESG news and reports leverage LLMs for fact-based policy analysis and media verification.
Despite these promising directions, existing efforts lack benchmarks that address the unique challenges of long-form, multimodal ESG documents. MMESGBench fills this gap by enabling document-level reasoning and cross-modal question answering, with direct implications for ESG disclosure validation, compliance assessment, and interactive stakeholder-orientated QA systems.

\section{MMESGBench}
We introduce MMESGBench, a benchmark tailored to the complexity of real-world ESG documents, capturing long-document reasoning, multimodal comprehension, and standard-driven semantics across diverse sources. MMESGBench is constructed via a structured pipeline consisting of model-driven generation with human-in-the-loop refinement to produce a high-fidelity, evaluation-ready benchmark, as delineated in Figure~\ref{fig:pipeline}.

\subsection{Document Collection}
To comprehensively reflect the diversity of ESG reporting practices, we curated documents across three primary categories aligned with the structure of real-world ESG disclosure ecosystems: 1)~\textit{Corporate ESG Reports}, including annual sustainability reports and CDP climate responses issued by companies; 2)~\textit{ESG Standards and Frameworks}, comprising regulatory and guidelines across four key subdomains: {Environment} (\eg GHG Protocol~\cite{ghgprotocol_standards_guidance}, TCFD~\cite{fsb2017tcfd}, ISO 14001~\cite{iso_standards}), {Social} (\eg UNGC, ISO 26000, SA8000~\cite{sa8000_standard}), {Governance} (\eg OECD Guidelines~\cite{oecd_standards}, ISO 37000), and {comprehensive standards} that span across multiple ESG dimensions (\eg GRI~\cite{gri2023standards}, SASB~\cite{sasb2023standards}, and TNFD); and 3)~\textit{Government and International Organization Documents}, which include global policy and regulatory frameworks such as the UN Sustainable Development Goals (SDGs), IPCC climate reports, and NGFS guidelines. This composition captures the corporate, regulatory, and policy dimensions of ESG content. 

From these categories, we select 45 representative and authoritative ESG documents based on the following criteria: (i) \textit{Multimodal richness}, ensuring the presence of textual content, tables, figures, chart, and layout-dependent visual structures; and (ii) \textit{Document complexity}, prioritizing documents of substantial length and structural depth, ranging from a few dozen to over 2,000 pages (average 157 pages). All documents are sourced from public repositories and retained in PDF format to preserve original layout and visual semantics. An overview of document types and statistics are presented in Figure~\ref{fig:dataset-overview}(b) and (c).

\subsection{Question-answer Generation}
The QA generation module constructs high-quality and diverse QA pairs grounded in multimodal ESG documents by leveraging multimodal LLMs and semantic clustering techniques.

For single-page QA generation, each document page is rendered as an image and processed by Qwen-VL-max~\cite{Qwen2VL}, a state-of-the-art vision-language model that has demonstrated strong performance on multimodal reasoning. We employ an example-based prompting strategy, where a small set of representative QA pairs are included to guide the model to generate contextually grounded questions. The model jointly attends to the text, layout, and visual elements of the input page and outputs candidate QA pairs. The generated questions fall into three major categories: (i) \textit{Factual extraction}, which retrieves explicit information from text or tables (\eg “What are the total Scope 3 emissions?”); (ii) \textit{Analytical or compliance-related}, which requires interpretation of ESG standards, regulatory logic, or metric synthesis (\eg “Does the GHG Protocol require companies to disclose their emission reduction targets?”); and (iii) \textit{Quantitative reasoning}, which involves arithmetic operations or comparisons over visual and textual elements, such as year-on-year trends or score aggregations. We also include visually grounded questions requiring table/chart/image understanding, intended to evaluate model performance on layout-aware reasoning.

For cross-page QA generation, we adopt a semantic clustering strategy built upon dense page-level embeddings and vector-based similarity retrieval. Specifically, each page is first encoded using PaliGemma-3B~\cite{beyer2024paligemma}, which integrates patch-wise visual encoding with token-level semantic fusion through Gemma-2B~\cite{team2024gemma}. The resulting embeddings are projected into a 128-dimensional space to obtain compact semantic representations for each page. To facilitate scalable similarity computation across long documents, we construct a FAISS-based vector index library~\cite{douze2024faiss} over all page embeddings. Nearest-neighbor retrieval is used to compute inter-page similarities, followed by clustering in the embedding space to form semantically coherent page groups, typically centered around shared topics such as emissions disclosures or governance practices. For each semantically coherent group, multimodal LLMs is prompted to generate QA pairs that require multi-page reasoning. These questions typically involve (i) \textit{aggregation across sections} (\eg “Summarize all the categories of Scope 3 GHG emissions disclosed across the report.”), (ii) \textit{temporal or metric comparison} (\eg “How much faster did sea level rise between 2006–2018 compared to 1901–1971?”), and (iii) \textit{causal linkage or referential reasoning} (\eg “What are the potential risks and impacts of enhanced weathering?”). This semantic clustering approach enables the discovery of latent document structures and supports scalable generation of cross-page QA pairs grounded in multi-hop ESG reasoning tasks.

To assess model robustness, we also include unanswerable QA pairs, generated at the document- or section-level by prompting the model to produce plausible yet unsupported questions aligned with the document’s theme, while ensuring no corresponding evidence exists. These instances are essential for benchmarking answerability detection and hallucination resistance in LLMs.
All QA pairs are annotated with its originating modality (text, table, chart, image, layout), type (single-page, cross-page, unanswerable), supporting detailed analysis of layout reasoning, multi-hop inference, and model reliability in ESG contexts.

\begin{table*}[ht]
\centering
\small
\caption{Evaluation of various models on MMESGBench (the best results are in \textbf{bold}, and the second best ones are \underline{underlined})}
\resizebox{\textwidth}{!}{%
\begin{tabular}{l c c c c c c c c c c c}
\toprule
& & \multicolumn{5}{c}{\textbf{Evidence Modalities}} &\multicolumn{3}{c}{\textbf{Evidence Location}} & \multicolumn{2}{c}{\textbf{Overall}} \\
\cmidrule(lr){3-7} \cmidrule(lr){8-10} \cmidrule(lr){11-12}
\textbf{Method} & \textbf{\# Pages} & \textbf{TXT} & \textbf{LAY}& \textbf{CHA} & \textbf{TAB} & \textbf{IMG} & \textbf{SIN} & \textbf{MUL} & \textbf{UNA} & \textbf{ACC} & \textbf{F1} \\
\midrule
\multicolumn{10}{c}{\textbf{Text Pipeline}} \\
\midrule
\textcolor{gray}{\textit{LLMs}} \\
ChatGLM-128k           & up to 120 & 10.0 & 9.3 & 6.3 &  6.2&  11.8 & 8.5 &  10.4 & 41.9 & 14.3 & 9.6 \\
Mistral-Instruct-v0.1  & up to 120 & 11.0 & 12.6 & 5.3 &  7.4 &  13.3 & 10.4 &  10.6 &  \textbf{84.5} & 22.2 & 14.2 \\
Qwen-14B-Chat  & up to 120 &10.5  &10.7  &5.8  &7.7  &9.5   &9.4  &10.3   &77.7   &20.4  &12.9  \\
Deepseek-llm-7b-chat & up to 120 & 7.8 & 8.1 & 1.8 &  4.0 &  7.1 & 6.0 &  9.2 &  \underline{79.1} & 18.3 & 9.8 \\
Qwen Max & up to 120 & 27.6 & 30.5 & 27.7 &  26.5 &  32.3 & 26.0 & 30.0 &  10.1 & 24.5 & 25.1 \\
\textcolor{gray}{\textit{Text RAG}} \\
ColBERT + Mistral-Instruct      &/          &40.1  &30.3  &24.0  &36.1  &33.9   &31.4  &32.8  &63.5  &37.0  &35.9  \\
ColBERT + Qwen Max     &/         &48.4  &40.8  &32.8  &\underline{45.0}   &40.2   &46.3  &40.4  &47.9  &41.5  & 40.9 \\
\midrule
\multicolumn{10}{c}{\textbf{Multi-modal Pipeline}} \\
\midrule
\textcolor{gray}{\textit{Multi-modal LLMs}} \\
DeepSeek-VL-Chat       & up to 120   &10.8  &7.4   &7.8   &8.6  &13.3  &10.2  &9.8  &42.6 &15.2  &10.4 \\
MiniCPM-Llama3-V2.5      & up to 120 &15.4  &12.7  &10.0  &11.4   &17.2   &14.9  &12.7  &9.5  &13.5  &13.2 \\
InternLM-XC2-4KHD            & up to 120 &12.9  &12.0  &8.3  &9.4   &14.8   &13.0  &9.9  &24.3  &14.0  &11.6 \\
Qwen2-VL-7B           & up to 120         &18.0  &16.4  &12.0  &14.6   &24.2   &17.0  &18.0  &33.1  &19.8  &17.1 \\
InternVL-Chat-V1.5            & up to 120         &13.3  &13.9  &3.5  &8.0 &11.0   &12.4   &10.4  &49.3  &17.8  &12.4\\
Qwen-VL-Max           & up to 120         &42.6  &40.5  & 36.5 &41.2   & 41.2  &38.2  &43.0  &41.9  &40.0  &38.3 \\

\textcolor{gray}{\textit{Multi-modal RAG}} \\
ColPali+Qwen2-VL 7B              & 1         &26.4  &24.9  &17.2  &25.0   &26.1   &24.7  &23.8  &51.4  &28.7  &25.4 \\
ColPali+Qwen2-VL 7B              & 5         &30.0  &28.1  &22.5  &28.1   &36.0   &27.0  &30.1  &48.6  &31.4  &28.9 \\


ColPali+Qwen-VL Max              & 1         &\underline{50.8}  &\underline{43.2}  &\underline{41.5}  &42.5   &\underline{51.9}   &\underline{48.1}  &\underline{42.5}  &57.5  &\underline{48.2}  &\underline{47.1} \\
ColPali+Qwen-VL Max   &5   &\textbf{55.5}  &\textbf{49.4}  &\textbf{48.0}  &\textbf{52.3}   &\textbf{57.2}   &\textbf{52.9}  &\textbf{49.2}  &52.0  &\textbf{51.8}  &\textbf{51.3} \\
\bottomrule
\label{table:res}
\end{tabular}}
\end{table*}

\subsection{Quality Control}
To ensure high annotation fidelity, we implement a two-stage quality control pipeline that combines LLM-based verification with expert-in-the-loop calibration. First, we filter out weakly grounded questions by applying no-context inference using Qwen-Max; questions that can be confidently answered without access to the document are removed. For each remaining QA pair, we re-infer the answer by providing the evidence page and question to multimodal LLM. The model's predicted answer is then compared with the annotated reference, and samples with low token-level F1 score or semantic mismatch are flagged. In parallel, the model evaluates each QA pair along the following three dimensions: (i) \textit{accuracy}, which reflects alignment between the answer and supporting evidence; (ii) \textit{completeness}, which considers coverage of all relevant content including text, tables, and layout-dependent elements; and (iii) \textit{answerability}, which assesses whether sufficient evidence exists on the given page to support the answer. Besides, each QA pair is further assigned a difficulty score by LLM based on the depth of reasoning required and the complexity of involved modalities.

Flagged examples are subsequently reviewed by experts and trained annotators via a retrieval-assisted interface. Reviewers assess evidence traceability, factual consistency, and multimodal alignment, making revisions or discarding low-quality samples as needed. This layered validation process ensures that MMESGBench provides a reliable foundation for evaluating multimodal LLMs in complex, document-centric ESG reasoning scenarios.

\subsection{Dataset Overview}
MMESGBench consists of 933 QA pairs constructed from 45 long-form ESG documents. This dataset spans across a diverse set of ESG document types with substantial variation in length, structure, and source complexity (see Figure~\ref{fig:dataset-overview}(b) and (c)). The QA pairs are distributed across three reasoning scopes: single-page (58.5\%), cross-page (25.6\%), and unanswerable (15.9\%), enabling evaluation across localized, multi-hop, and adversarial settings. Each QA instance is annotated with fine-grained modality tags, covering text, tables, charts, layout, and images, reflecting the heterogeneous nature of ESG documents. Evidence sources comprise both structured and unstructured elements, with pure text and layout accounting for the largest share. Answer formats include strings, numbers, and lists, supporting both factual and quantitative reasoning, detailed statistic shown in Figure~\ref{fig:dataset-overview}(a). Figure~\ref{fig:dataset-overview}(d) visualizes the distribution of QA evidence across document positions, confirming broad coverage across structural regions. Together, these characteristics make MMESGBench a comprehensive and realistic benchmark for evaluating multimodal reasoning in complex ESG reporting contexts.

\subsection{Potential Applications}
MMESGBench is designed for the ESG domain, where long-form, multimodal documents like sustainability reports and regulatory disclosures underpin transparency, compliance, and investment decisions. It supports downstream tasks such as automated ESG report validation, quantitative metric extraction, disclosure alignment with standards such as GRI, SASB, as well as climate risk analysis assessment. It also enables interactive use cases such as stakeholder-orientated QA systems, ESG policy assistants, and sustainability chatbots that can respond to complex, evidence-grounded queries. These rely on reasoning across text, tables, visuals, and layouts—all covered by MMESGBench.

MMESGBench also serves as a testbed for advancing LLM and multimodal research, especially in long-document understanding and RAG. It enables evaluation of techniques for evidence retrieval, multimodal fusion, context selection, and hallucination mitigation. It offers a unified benchmark for multimodal grounding, layout-aware reasoning, and multi-hop inference, supporting comparison across general and domain-specific models.


\section{Evaluation and Analysis}
\subsection{Evaluation Protocol and Models}
Following recent multimodal document understanding benchmarks such as MMLongBench~\cite{ma2025mmlongbench}, we adopt a three-stage evaluation protocol, consisting of free-form response generation, automatic short-form answer extraction, and rule-based score computation. This design ensures consistent evaluation across models with diverse decoding styles while emphasizing document-level reasoning capabilities. We report two key metrics: \textit{answer accuracy}, measuring exact match with reference answers, and {generalized macro-F1}, which balances performance across answerable and unanswerable questions by accounting for partial matches and abstentions. We further provide performance breakdowns by evidence modality (\eg text, layout, chart, table, and image) and evidence location (single-page, cross-page, unanswerable), enabling fine-grained analysis of model behavior across diverse conditions.

We evaluate 15 models across three categories: text-only LLMs, multimodal LLMs, and RAG pipelines. For text-only models, we use OCR to extract text and truncate or segment it based on context limits. This group includes ChatGLM-128k~\cite{glm2024chatglm}, Mistral-Instruct-v0.1~\cite{jiang2023mistral7b}, Qwen-14B-Chat~\cite{bai2023qwen}, DeepSeek-7B-Chat~\cite{bi2024deepseek}, and Qwen-Max accessed via API. These models are capable of long-range textual reasoning but do not process visual or structural content. Multimodal LLMs are evaluated by rendering document pages as images and concatenating them based on model capacity. We include open-source models such as DeepSeek-VL~\cite{lu2024deepseek}, MiniCPM-V2.5~\cite{yao2024minicpmv}, InternLM-XC2~\cite{dong2024internlm}, InternVL~\cite{chen2024internvl}, and Qwen-VL-7B~\cite{Qwen2VL}. We also include Qwen-VL-Max, a proprietary model with extended visual context support, as a strong baseline. Retrieval-augmented models are especially relevant for MMESGBench given its document length (157 pages on average) and sparse evidence distribution. We use retrievers ColBERT~\cite{khattab2020colbert} for text-only models and ColPali~\cite{faysse2024colpali} for multimodal LLMs to identify relevant content chunks or pages. These retrieved segments are then passed to either text-based or multimodal decoders. This setup reflects practical and scalable workflows for QA over long-form ESG documents.

\subsection{Main Results and Findings}
Table~\ref{table:res} summarizes the performance of various models across different evidence modalities, reasoning types, and overall accuracy. It is obvious that multimodal models substantially outperform their text-only counterparts. For instance, Qwen-VL-Max achieves significantly stronger results than Qwen-Max, yielding over 60\% relative improvement in accuracy, with especially notable gains on layout and chart-based questions. Retrieval-augmented models further enhance performance. Compared to its non-retrieval variant, ColPali+Qwen-VL-Max (5 pages retrieval setting) improves the overall accuracy by 30\% and layout-specific performance by 15\%. These gains further highlight the importance of targeted evidence selection in long documents, where full-context processing is infeasible and irrelevant content can hinder reasoning.

Despite strong overall improvements, key challenges persist. Most models still underperform on chart-based questions, indicating unresolved limitations in fine-grained spatial and numerical reasoning. Layout-intensive cases also expose weaknesses in structure-aware modeling, especially for models without visual encoders or retrieval module. Additionally, smaller text-only models tend to over-predict unanswerable cases, resulting in inflated accuracy on negative samples but lower overall F1 score. These trends underscore the importance of both visual grounding and targeted evidence retrieval for robust ESG document understanding. These findings reaffirm that accurate ESG QA requires both visual-semantic fusion and precise evidence localization, and demonstrate how MMESGBench surfaces failure modes that are easily overlooked in conventional benchmarks.

\begin{figure}[t]
    \centering
    \begin{subfigure}[t]{0.53\linewidth}
        \centering
        \includegraphics[width=\linewidth]{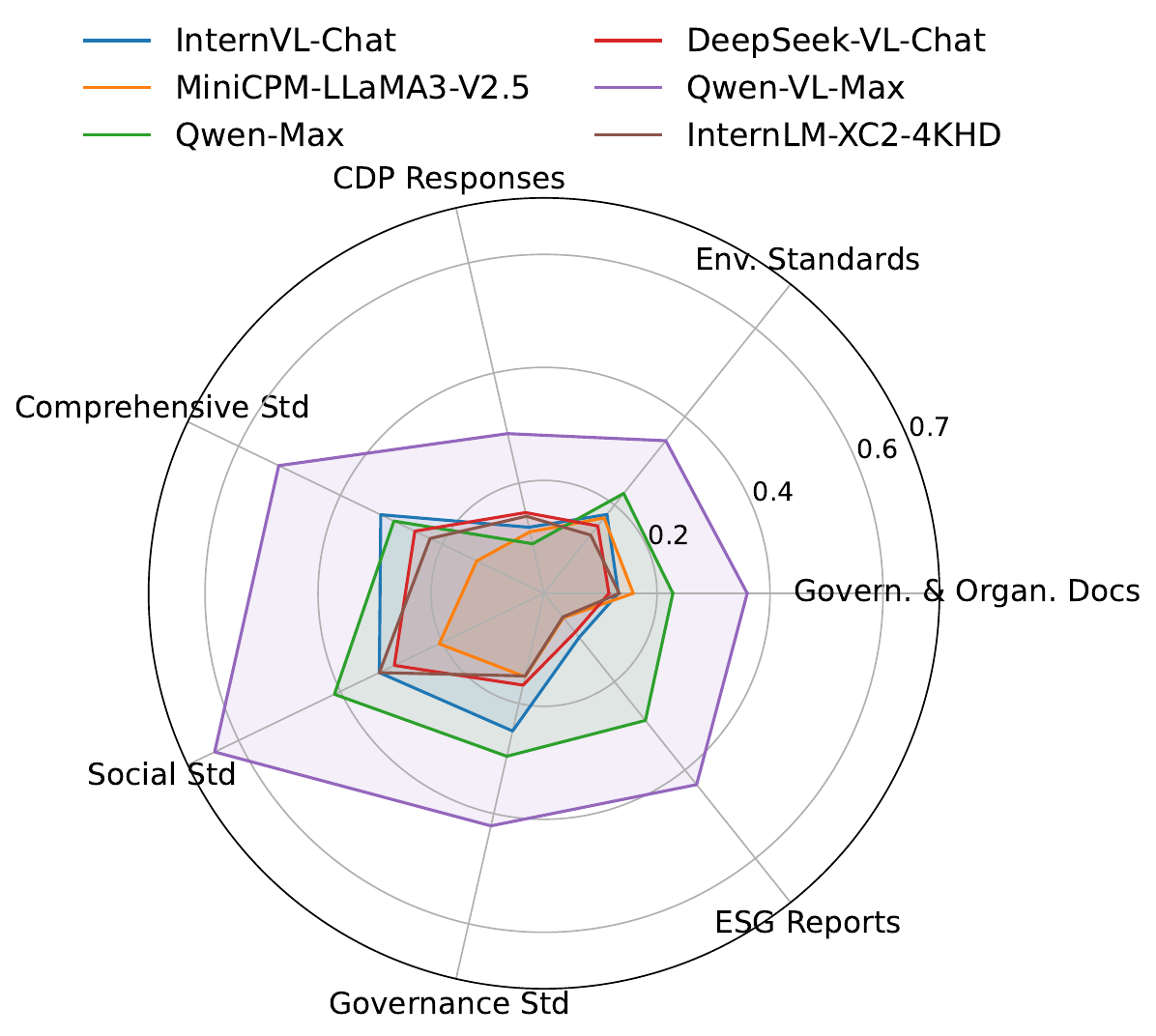}
        \caption{Document Type}
        \label{fig:doc_type_radar}
    \end{subfigure}
    \hfill
    \begin{subfigure}[t]{0.45\linewidth}
        \centering
        \includegraphics[width=\linewidth]{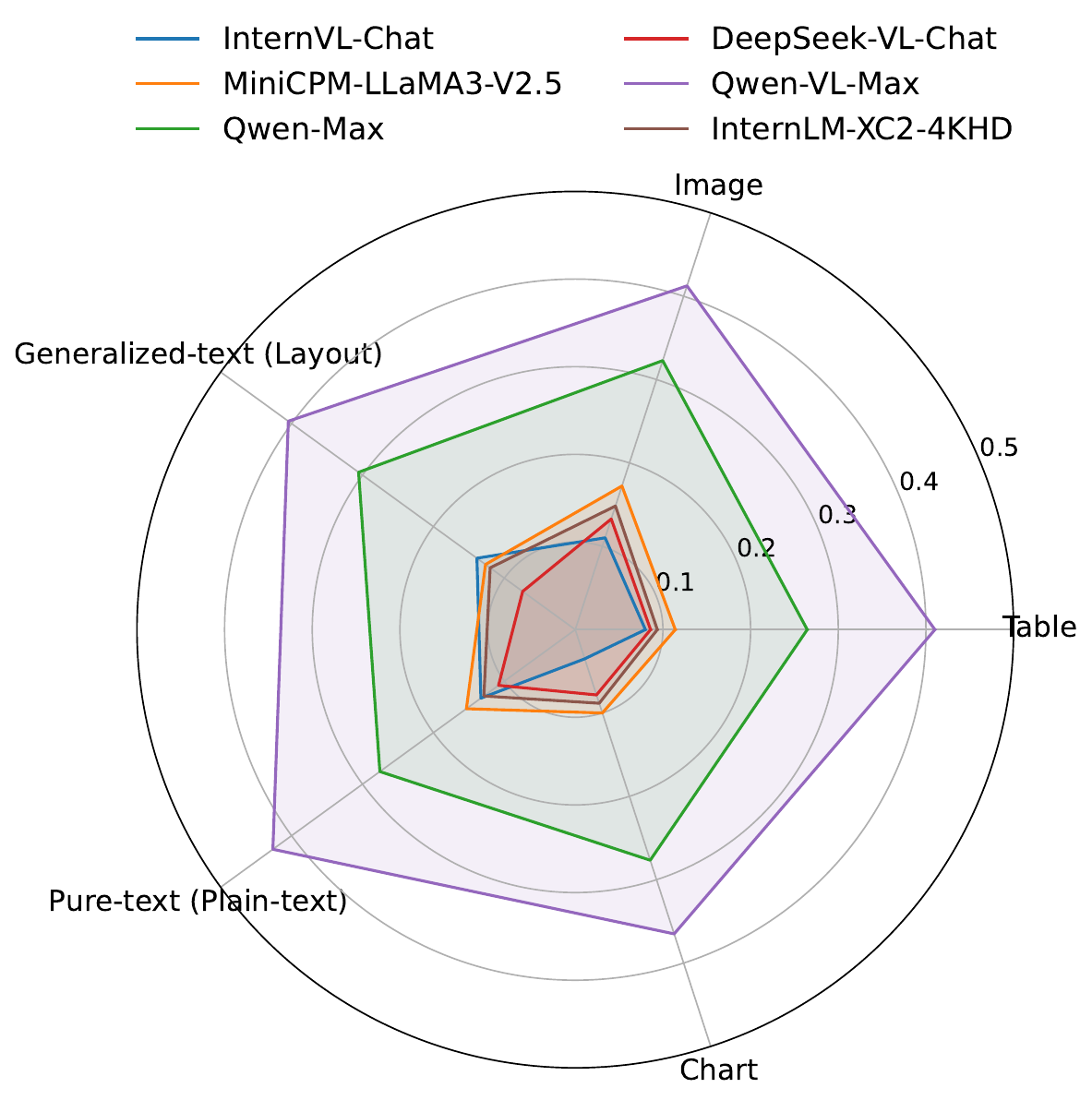}
        \caption{Evidence Source}
        \label{fig:evidence_type}
    \end{subfigure}
    \caption{Comparative analysis of multimodal QA model performance across document types and evidence modalities.}
    \label{fig:radar}
\end{figure}

Figure~\ref{fig:radar} shows performance variations across document types and evidence modalities. Overall, Qwen-VL-Max achieves the highest and most consistent results overall, particularly excelling on complex documents such as ESG reports, social standards, and comprehensive frameworks. In terms of evidence modality, models perform well on plain text and layout-based questions, which offer clearer semantic cues. However, accuracy declines significantly for tables, images, and especially charts, suggesting that fine-grained visual reasoning and numerical interpretation remain major challenges for current multimodal models.

\section{Conclusion}
We present MMESGBench, a novel benchmark for long-form, multimodal document understanding in the ESG domain. 
This dataset captures the structural and semantic complexity of real-world ESG disclosures, encompassing diverse document types, evidence modalities, and reasoning challenges.  Built through a quality-controlled, human-AI collaborative pipeline, MMESGBench supports comprehensive assessment of LLMs, multimodal LLMs, and retrieval-augmented systems. It provides a practical testbed for ESG applications such as disclosure validation and compliance analysis, while also advancing research in multimodal reasoning and long-context retrieval. Going forward, we plan to extend the benchmark with richer annotation schemes, support for generative tasks, and more fine-grained evaluation protocols tailored to real-world ESG decision workflows.

\section{Acknowledgments}
This research is partially supported by the RIE2025 Industry Alignment Fund – Industry Collaboration Projects (IAF-ICP) (Award I2301E0026), administered by A*STAR, as well as supported by Alibaba Group and NTU Singapore through Alibaba-NTU Global e-Sustainability CorpLab (ANGEL).
It is also supported by the Joint NTU-UBC Research Centre of Excellence in Active Living for the Elderly (LILY) and Jinan–NTU Green Technology Research Institute (GreenTRI).

\bibliographystyle{ACM-Reference-Format}
\balance
\bibliography{sample-base}

\end{document}